\documentstyle[epsfig]{mn}

\title[Accretion onto low-mass Pop III stars]{Suppression of accretion onto low-mass Population III stars}

\author[J.L. Johnson and S. Khochfar]
{Jarrett L. Johnson\thanks{E-mail: jjohnson@mpe.mpg.de} and Sadegh Khochfar \\
Max-Planck-Institut f{\"u}r extraterrestrische Physik, 
Giessenbachstra\ss{}e, 85748 Garching, Germany \\
Theoretical Modeling of Cosmic Structures Group \\}

\pubyear{2010}

\begin{document}
\maketitle
\topmargin-1cm

\begin{abstract}
Motivated by recent theoretical work suggesting that a substantial fraction of Population (Pop) III 
stars may have had masses low enough for them to survive to the present day, we consider the
role that the accretion of metal-enriched gas may have had in altering their surface composition, thereby disguising them as Pop~II stars.
We demonstrate that if weak, Solar-like winds are launched from low-mass Pop~III stars formed in the 
progenitors of the dark matter halo of the Galaxy, then such stars are likely to avoid significant enrichment 
via accretion of material from the interstellar medium.  We find that at early times accretion is easily prevented 
if the stars are ejected from the central regions of the haloes in which they form, either by dynamical interactions with more massive Pop~III
stars, or by violent relaxation during halo mergers.  While accretion may still take place during passage through sufficiently dense 
molecular clouds at later times, we find that the probability of such a passage is generally low ($\la$ 0.1), 
assuming that stars have velocities of order the maximum circular velocity of their host haloes and accounting for 
the orbital decay of merging haloes.  In turn, due to the higher gas density required for accretion onto stars with higher velocities, we find an
even lower probability of accretion ($\sim$ 10$^{-2}$) for the subset of Pop~III stars formed at $z$ $>$ 10, 
which are more quickly incorporated into massive haloes than stars formed at lower redshift. 
While there is no a priori reason to assume that low-mass Pop~III stars do not have Solar-like winds, 
without them surface enrichment via accretion is likely to be inevitable.  We briefly discuss 
the implications that our results hold for stellar archaeology.
\end{abstract}


\section{Introduction}
Knowledge of the initial mass function (IMF) of the first stars is critical to our understanding of the earliest stages of galaxy formation, 
one of the key frontiers of modern cosmology (see e.g. Barkana \& Loeb 2001; Ciardi \& Ferrara 2005; Bromm et al. 2009; Benson 2010).   
While the Population~(Pop)~III IMF is widely believed to have been top-heavy due to the limited cooling properties of primordial gas 
(see e.g. Bromm \& Larson 2004), various scenarios for the formation of low-mass primordial stars have been discussed for some time 
(e.g. Palla et al. 1983; Nakamura \& Umemura 2001; Omukai \& Yoshii 2003; Salvaterra et al. 2004; Machida et al. 2005; Silk \& Langer 2006).  
Most recently, high-resolution simulations of Pop~III star formation have challenged the view that the first stars were very massive single objects 
formed in isolation (e.g. Abel et al. 2002; Bromm et al. 2002), instead suggesting that many formed as binaries (Turk et al. 2009; 
Stacy et al. 2010) or in clusters of lower mass stars (Clark et al. 2010; Greif et al. 2010a).   
These results also raise a challenge to the postulated existence of a critical metallicity, 
below which stars can only form with a top-heavy IMF 
(e.g. Bromm \& Loeb 2003; Santoro \& Shull 2006; Schneider et al. 2006; Frebel et al. 2007; but see also Jappsen et al. 2009).

Further motivation to consider the formation and fate of low-mass Pop~III stars comes from the fact that no 
low-metallicity star has yet been found to exhibit the chemical signature of pair-instability supernovae (SNe), one of 
the primary predicted end states of very massive Pop~III stars (see e.g. Fryer et al. 2001; Heger et al. 2003).  While this observation in itself 
does not imply that low-mass Pop~III stars formed, it is consistent with a Pop~III IMF that is not dominated 
by very massive stars (see Karlsson et al. 2008); instead, the observed chemical abundances of low-metallicity stars 
(see Beers \& Christlieb 2005; Frebel 2010) suggest that they formed from material enriched by the explosions of 
15-40 M$_{\odot}$ Pop~III stars (e.g. Umeda \& Nomoto 2003; Iwamoto et al. 2005; Joggerst et al. 2010).  
Altogether, there remains the distinct possibility that some 
fraction of Pop~III stars had masses low enough ($\la$ 0.8 M$_{\odot}$) that they may have survived to the present day. 
In turn, this raises once again the long-standing question (e.g. Bond 1981): Where are such low-mass Pop~III stars today? 


One explanation for why Pop~III stars may not be identified in the Galaxy today is that they do not appear as such, their
surfaces having been enriched due to the accretion of heavy elements from the interstellar medium (ISM) during the course of their long lives. 
Enrichment by accretion has been studied by numerous authors, in the context of the present-day Milky Way 
(e.g. Mestel 1954; Talbot \& Newman 1977; Alcock \& Illarionov 1980; Yoshii 1981; Iben 1983; Frebel et al. 2009).  Using recently obtained
kinematic information on a large number of metal-poor stars, Frebel et al. (2009) modelled the enrichment of these stars individually
over the past $\sim$ 10 Gyr, finding that in almost all cases the amount of metals accreted onto the stars was far below what would be
needed to explain their observed surface metallicities.  Based on this result, it was concluded that the accretion of enriched material during
passage through the disk of the Galaxy has not significantly altered the surface composition of observed low-metallicity stars.  

However, as the first stars formed during the earliest stages of the assembly of the Milky Way, it is important to 
also consider the degree to which the accretion of enriched material may have altered their surface compositions during this entire assembly process.
Recently, this has been studied by Komiya et al. (2009, 2010) using the extended Press-Schechter formalism to model the hierarchical growth of the 
Milky Way.  Their findings suggest that significant accretion of enriched material occurs soon after the 
formation of a Pop~III star, before its host minihalo undergoes a merger.  

Here we present a complementary study in which we evaluate the likelihood that low-mass Pop~III stars may avoid enrichment via accretion from the ISM,  
using a simple model for their dynamical evolution during the hierarchical assembly of the Galaxy.  In particular, we investigate the impact that a weak, 
Solar-like stellar wind may have in limiting accretion, as it is perhaps only by such a wind that accretion onto a low-mass star
can be completely prevented.

In the next two Sections, we review the criteria for the suppression of accretion by a stellar wind and 
we describe our model for the dynamical evolution of low-mass Pop~III stars. 
In Section 4, we present our findings on the conditions for, and the likelihood of, enrichment by accretion from the ISM in the assembly of the Galaxy.  
Finally, we briefly discuss the implications of our findings and give our conclusions in Section 5.

\section{Suppression of accretion by a Solar-like stellar wind}
Just as in the case of the Sun, the outer envelope of a low-mass Pop~III star is expected to be convective (e.g. Yoshii 1981; Fujimoto et al. 1995).  
Therefore, as in the Sun, this convection may drive a dynamo which produces a magnetic field emanating from the surface of the star.  In turn, 
as the corona of the Sun is likely generated by processes associated with its magnetic field (e.g. Hassler et al. 1999),
such a star may also have a hot corona from which a weak, Solar-like wind is launched.  

To evaluate the effect that such a stellar wind would have in suppressing the accretion of material from the ISM, we follow Talbot \& Newman (1977) and
define a minimum density $n_{\rm min}$ of the ISM required for accretion to take place when a stellar wind is launched with the same 
density and speed as the Solar wind:

\begin{equation}
n_{\rm min} =  0.8 {\rm cm}^{-3} \left(v^{2}_{\rm *} + c^{2}_{\rm s}\right)   \mbox{\ ,}
\end{equation}
where $v_{\rm *}$ is the velocity of the star with respect to the ISM and $c_{\rm s}$ is the sound speed of the gas, both in units of km s$^{-1}$.  
Where in its orbit the velocity of a star is large enough that $n_{\rm min}$ exceeds the 
density $n_{\rm gas}$ of the ISM through which the star passes, then the accretion of gas from the ISM is prevented.

For simplicity, in the following we shall adopt the formula for $n_{\rm min}$ given by Talbot \& Newman (1977) directly, although it is 
derived for the specific case of the Sun.  However, the radius of a Pop~III star with mass $\sim$ 0.8 M$_{\odot}$, which may still 
be on the main-sequence today, is expected to be comparable to the Solar radius 
(e.g. Weiss et al. 2000; Suda et al. 2007).  Thus, we expect that the velocity of a wind from such stars, 
roughly the escape velocity, will also be similar to the Solar value.  In turn, if the density of the wind is also comparable to 
that of the Solar wind, then the above formula is appropriate.  Lacking detailed knowledge of such winds from Pop~III stars, 
we limit ourselves to considering the single case of a Solar-like wind.   


\section{Modelling Stellar dynamics in the assembly of the Halo}
Here we consider the nature of stellar orbits within the progenitors of the dark matter (DM) halo of the Galaxy, 
in order to estimate the typical velocities of low-mass Pop~III stars with respect to the ISM.
As the detailed properties of the high-redshift progenitor haloes of the Milky Way are unknown, 
we begin by adopting the following simple form for the gravitational potential, consistent with
a $\rho$ $\propto$ $r^{-2}$ matter density profile (White 1985), as well as with the observed 
velocity curve of the Galaxy (Xue et al. 2008) and with cosmological simulations of the formation of the first protogalaxies 
(e.g. Johnson et al. 2007; Wise \& Abel 2008):

\begin{equation}
\Psi(r) = v_{\rm max}^2 {\rm ln}\left(\frac{r}{r_{\rm 0}}\right) \mbox{\ .}
\end{equation}
Here $r_{\rm 0}$ is arbitrary and $v_{\rm max}$ is the maximum circular velocity of a halo of mass $M_{\rm h}$, given by  
(e.g. Barkana \& Loeb 2001)

\begin{equation}
v_{\rm max} = \left(\frac{{\rm G} M_{\rm h}}{r_{\rm vir}} \right)^{\frac{1}{2}} \sim 20 {\rm km} {\rm s^{-1}} \left(\frac{M_{\rm h}}{10^8 {\rm M_{\odot}}}\right)^{\frac{1}{3}} \left(\frac{1+z}{10}\right)^{\frac{1}{2}} \mbox{\ .}
\end{equation}
A stellar orbit can then be defined by an apogalactic distance $r_{\rm ap}$ and an eccentricity $e$, given by

\begin{equation}
e = \frac{r_{\rm ap} - r_{\rm pr}}{r_{\rm ap} + r_{\rm pr}} \mbox{\ ,}
\end{equation}
where $r_{\rm pr}$ is the perigalactic distance of the orbit (e.g. Chiba \& Beers 2000).
With these parameters, the specific orbital angular momentum $l$ of the star is given by

\begin{equation}
l^2 = r^2_{\rm ap} v^2_{\rm max} \frac{(1-e)^2}{2e}{\rm ln}\left(\frac{1+e}{1-e}\right)  \mbox{\ ,}
\end{equation}
and the velocity of the star at radius $r$ is 

\begin{equation}
v^2_{\rm *}(r) = v_{\rm max}^2 \left({\rm ln}\left(\frac{r_{\rm ap}}{r}\right) +  \frac{(1-e)^2}{2e}{\rm ln}\left(\frac{1+e}{1-e}\right) \right) \mbox{\ .} 
\end{equation}
We note that even for a very eccentric orbit with $e$ = 0.99,  the velocity $v^2_{\rm *}$ of the star changes by only a factor 
ln($r_{\rm ap}$/$r$) $\le$ ln((1+$e$)/(1-$e$)) $\sim$ 5 throughout its orbit; hence, to within a factor of order unity, $v_{\rm *}$ $\sim$ $v_{\rm max}$.

Strictly speaking, the velocity $v_{\rm *}$ we derive here is the velocity of the star relative to the center of its host dark matter halo.  
However, for simplicity, we shall treat this as the velocity of the star relative to the ISM.  
While this assumption is clearly not valid for the case of stars formed, for example, in the disk of the 
present-day Milky Way (see e.g. Talbot \& Newman 1977; Yoshii 1981), it is more tenable for Pop~III stars 
formed at high redshift, as the orbits of such stars would be altered numerous times in the frequent mergers that their host haloes undergo in the
assembly of the Galaxy.  Thus, we expect that any extant low-mass Pop~III stars are likely to have orbits that are in general not oriented 
in a particular direction with respect to the motion of the ISM (see e.g. Carollo et al. 2007), in which case a {\it typical} such star is indeed likely to have 
a velocity relative to the ISM that is comparable to $v_{\rm *}$ $\sim$ $v_{\rm max}$.     


Furthermore, in the calculations that follow, we shall neglect the possibility that intracloud motions are of the order of $v_{\rm max}$ and thereby impact the 
value of $n_{\rm min}$ within a cloud, as cosmological simulations of the formation of high redshift dwarf 
galaxies (Wise \& Abel 2007a; Greif et al. 2008) show that the velocity 
dispersion within clouds with $n_{\rm gas}$ $\ga$ $n_{\rm min}$ are well below $v_{\rm max}$.  
We note, however, that cloud collisions may lead to temporarily higher intracloud velocity dispersions.

\section{Probability of accretion onto a Pop~III star with a Solar-like wind}
Here we consider the two primary situations in which accretion onto a single low-mass star with a Solar-like wind may take place: 
from the dense gas at the center of the DM halo in which it forms and during passage through a dense gas cloud at a later time in the
hierarchical assembly of the Milky Way.  For the latter case, we consider separately the most massive progenitor halo 
of the Milky Way, in which the earliest Pop III stars likely formed (e.g. Brook et al. 2007), and the full population of progenitor
haloes which may or may not have hosted Pop~III stars, due to chemical and radiative feedback (e.g. Gao et al. 2010).

\subsection{Accretion of isothermal gas in the host halo}
For the density $n_{\rm gas}$ of the ISM in a DM halo hosting the formation of a Pop~III star at high redshift, 
here we consider the case of an isothermal density profile with the following scaling, roughly consistent with the properties of the gas in Pop~III star-forming haloes, as found in numerical simulations (e.g. O'Shea \& Norman 2007; Wise \& Abel 2007b):

\begin{equation}
n_{\rm gas} = n_{\rm gas, vir} \left(\frac{r}{r_{\rm vir}}\right)^{-2} \mbox{\ ,}  
\end{equation}
where the number density of baryons at the virial radius $r_{\rm vir}$ is given by

\begin{equation}
n_{\rm gas,vir} \sim 178 \Omega_{\rm b}  (1+z)^3 \frac{\rho_{\rm crit}}{m_{\rm H}} \mbox{\ .} 
\end{equation}
Here $\Omega_{\rm b}$ is the ratio of the cosmic baryon density to the critical density $\rho_{\rm crit}$ of the Universe at redshift $z$ = 0, and $m_{\rm H}$ is the mass of the hydrogen atom.  To determine where in a halo of mass $M_{\rm h}$ a star with a Solar-like wind may accrete gas, we compare the gas density, given by equation (7), to the minimum density required for accretion, given by equation (1).  Taking, as discussed in Section 3, $v_{\rm *}$ = $v_{\rm max}$, and assuming a sound speed $c_{\rm s}$ $\la$ $v_{\rm max}$, we have that the condition for accretion (i.e. $n_{\rm gas}$ $\ge$ $n_{\rm min}$) is only satisfied if

\begin{equation}
\left(\frac{M_{\rm h}}{10^8 {\rm M_{\odot}}}\right)^{-\frac{2}{3}} \left(\frac{1+z}{10} \right)^2 \left(\frac{r}{r_{\rm vir}}\right)^{-2}
\ga 10^4 \mbox{\ .}  
\end{equation}
Simplifying this, we find that the dependences on the halo mass and redshift cancel out, leaving us with the following condition for accretion:

\begin{equation}
r \la 10 {\rm pc}.
\end{equation}
That is, for the case of an isothermal gas at the virial temperature of its host halo, a star will only accrete if
it is on an orbit that brings it within a physical distance of $\sim$ 10 pc of the center of the halo.

\begin{figure*}
\includegraphics[width=4.5in]{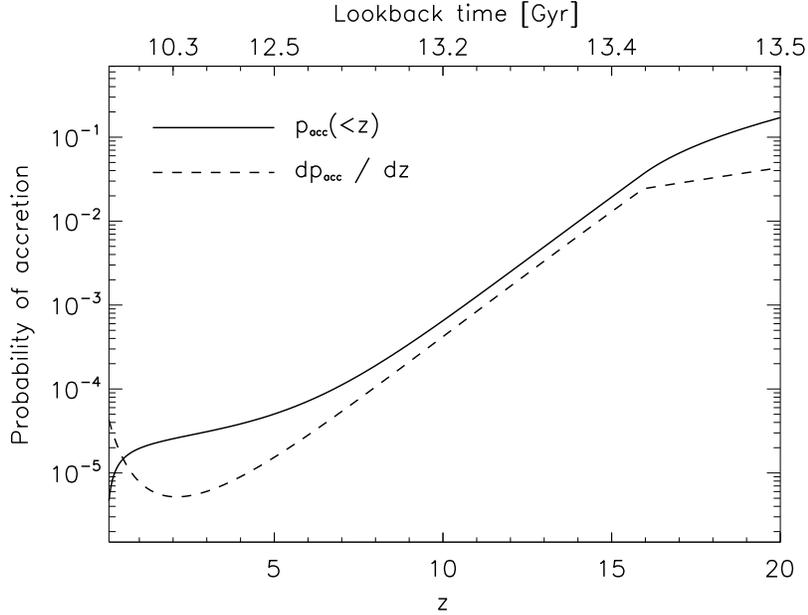}
\caption{The probability that a star with a Solar-like wind passes through a cloud with sufficiently high density that accretion 
occurs, within the main progenitor of the halo of the Milky Way.  The dashed line shows the probability of accretion per unit redshift $z$, 
while the solid line shows the probability for accretion taking place after redshift $z$.  
Due to the lower stellar velocities within the halo at earlier times, and to the larger volume-filling factor of the gas that can thus be accreted,
there is a much higher likelihood of accretion at higher redshift (see also Komiya et al. 2010 for the case of no stellar wind).  Note that the kink in the probability 
per unit redshift at $z$ $\sim$ 16 is due to the power-law break in the function adopted for $f_{\rm cloud}$ (see equation 14).  The
low probability of accretion ($p_{\rm acc}$ $\le$ 0.2) overall suggests that low-mass Pop~III stars with Solar-like winds may typically avoid accretion altogether.  
}
\end{figure*}

Given this condition, if dynamical interactions result in the ejection of 
a low-mass Pop~III star from the central $\sim$ 10 pc of the halo in which it forms, then the star will be prevented from accreting material.  
If the Pop~III IMF is top-heavy, albeit with some low-mass stars forming, then dynamical interactions during the formation of stellar clusters are likely to eject
the low-mass stars at higher velocities than in the case of present-day star formation (e.g. Greif et al. 2010a), which can be up to 
a few km s$^{-1}$ (e.g. Bate et al. 2003).  As such ejection velocities can be comparable to the virial velocity of the minihalo hosting Pop~III star formation, 
this could result in such low-mass stars being kicked out of the centers of their host haloes at early times.  Indeed, it may be the case that low-mass stars form
{\it because} they are kicked out of the center of the halo, the accretion of dense gas being halted before the star can grow to be massive 
(e.g. Bate et al. 2003; Clark et al. 2010).  Note that even for an ejection velocity of just 3-5 km s$^{-1}$, a low-mass Pop~III star would travel 
10 pc from the center of its host minihalo within the 2-3 Myr lifetime of the most massive stars that explode as SNe and enrich the ISM.  Therefore, low-mass
Pop~III stars may avoid accretion of metals from within their host haloes as a natural consequence of the process of their formation.

We also note that mergers of star-forming haloes, which for the minihalo progenitors of the first galaxies occur on timescales of 10$^{7}$ - 10$^8$ Myr 
(e.g. Greif et al. 2008; Gao et al. 2010), are likely to result in the violent relaxation of the stars in the halo, and this may also serve to 
kick low-mass stars to larger radii in the merged host halo. 

A further impediment to the accretion of metals is that this will not occur until the SNe ejecta from massive Pop~III stars 
re-collapses to high densities within the host halo, which generally occurs on a timescale of $\sim$ 10$^8$ yr
(see Wise \& Abel 2008; Greif et al. 2010b; but see also Whalen et al. 2008a).  Prior to this re-collapse, the ejecta will be at low density and may also 
be moving with high velocities, making accretion very unlikely.
If dynamical interactions within the initial stellar cluster or violent relaxation as a result of a merger, take place within this time, then a low-mass 
star is likely to avoid accretion in its host minihalo altogether.  

\subsection{Accretion from dense clouds}
While the condition we have just found dramatically limits the range of circumstances in which a low-mass Pop~III star can accrete metal-enriched gas within the halo 
in which it forms, it is still possible for accretion to take place if the star passes through a sufficiently dense molecular cloud in a more massive halo
at a later time.  Taking it that the star has a velocity relative to a dense cloud of $v_{\rm *}$ $\sim$ $v_{\rm max}$, and using equation (7) 
with the redshift dependence of $r_{\rm vir}$, we find the minimum density $n_{\rm min}$ of a cloud required for accretion to be 

\begin{equation}
n_{\rm min} \sim 500 {\rm cm}^{-3} \left(\frac{M_{\rm h}}{10^8 {\rm M_{\odot}}}\right)^{\frac{2}{3}} \left(\frac{1+z}{10}\right) \mbox{\ .}
\end{equation}
For a molecular cloud with with this density and with a characteristic size $r_{\rm cloud}$ $\sim$ 10 pc (e.g. Spitzer 1978), assuming the standard Bondi (1952) accretion rate, we find 
that this would result in the accretion of a total mass of $\sim$ 10$^{-7}$ M$_{\odot}$, leading to a stellar surface metallicity of 

\begin{eqnarray}
{Z}_{\rm surf} & \sim  & 10^{-7} {\rm Z_{\odot}}  \left(\frac{f_{\rm conv}}{10^{-3}} \right)^{-1}  \left(\frac{M_{\rm *}}{0.8 {\rm M_{\odot}}}\right)^2 \left(\frac{r_{\rm cloud}}{10 {\rm pc}}\right) \nonumber \\
& & \left(\frac{Z_{\rm met}}{{\rm 10^{-3} Z_{\odot}}}\right)  \left(\frac{M_{\rm h}}{10^8 {\rm M_{\odot}}}\right)^{-\frac{2}{3}} \left(\frac{1+z}{10}\right)^{-1} \mbox{\ ,}
\end{eqnarray}
where $Z_{\rm met}$ is the metallicity of the gas in the cloud, $M_{\rm *}$ $\la$ 0.8 M$_{\odot}$ is the mass of the star, and $f_{\rm conv}$ is the fraction of the stellar mass contained in the outer convective layer (e.g. Yoshii 1981).  

We have normalized this formula to the results of recent cosmological simulations of the earliest episodes of metal enrichment 
via powerful Pop~III SNe (Wise \& Abel 2008; Greif et al. 2010b), from which it has been found that the metallicity of the gas that re-collapses into haloes of mass $M_{\rm h}$ $\sim$ 10$^8$ M$_{\odot}$ at $z$ $\ga$ 10 is of the order of $Z_{\rm met}$ $\sim$ 10$^{-3}$ $Z_{\odot}$.  This metallicity is likely to be roughly an order of magnitude lower if the Pop~III IMF is not top-heavy (see e.g. Tumlinson 2010; Komiya et al. 2010).
For the case of the Milky Way at $z$ = 0, with a typical cloud size $r_{\rm cloud}$ $\sim$ 10 pc, $M_{\rm h}$ = 10$^{12}$ M$_{\odot}$, and $Z_{\rm met}$ = Z$_{\odot}$, we find $Z_{\rm surf}$ $\sim$ 10$^{-6}$ Z$_{\odot}$, in basic agreement with the estimate of the surface enrichment for the similar (but with $r_{\rm cloud}$ = 100 pc) case presented by Frebel et al. (2009).

In the following, we consider the probability of a star encountering a cloud of sufficient density that such enrichment via accretion may take place.

\subsubsection{Accretion within the main progenitor of the Galaxy}
An important specific case to consider is the probability of accretion onto low-mass Pop~III stars formed in the main progenitor halo of the Galaxy, as 
this is the most likely formation site of the earliest Pop~III stars, which formed before radiative or chemical feedback could begin to inhibit 
primordial star formation (e.g. Brook et al. 2007; Gao et al. 2010).

For this, we make use of the results from recent, high-resolution cosmological simulations of the formation of Milky Way-like haloes (Boylan-Kolchin et al. 2010).
These authors provide a best-fit formula for the mass assembly histories of their simulated haloes, as a function of redshift:

\begin{equation}
M_{\rm h}(z) = M_{\rm 0} (1+z)^{2.23} {\rm exp}\left[-4.9\left((1+z)^{\frac{1}{2}} - 1\right)\right] \mbox{\ ,}
\end{equation}
where we take the present-day mass of the halo to be $M_{\rm 0}$ = 10$^{12}$ M$_{\odot}$.
Comparing this formula with the results of Zhao et al. (2009) on the early growth of Milky Way-like haloes, we find good agreement up to at 
least $z$ $\sim$ 20 (see also Gao et al. 2010), at which point Pop~III star formation was likely ongoing (e.g. Bromm \& Larson 2004).  
Therefore, we shall use this formula for the growth history of the halo from $z$ = 20 to the present-day in our calculations.

Next, we shall assume the same density dependence of the volume-filling fraction $f_{\rm cloud}$($n_{\rm gas}$) of 
gas clouds with gas density $\ge$ $n_{\rm gas}$ as found by Talbot \& Newman (1977).  In particular, we shall take it that

\begin{equation}
f_{\rm cloud} = 5.6 \times 10^{-3} \left(\frac{n_{\rm gas}}{10^3 {\rm cm^{-3}}} \right)^{-\beta} \mbox{\ ,}
\end{equation}
where $\beta$ = 0.9 for $n_{\rm gas}$ $\le$ 10$^{3}$ cm$^{-3}$ and  $\beta$ = 3 for $n_{\rm gas}$ $>$ 10$^{3}$ cm$^{-3}$.  
We note that this value of $\beta$ for $n_{\rm gas}$ $>$ 10$^3$ cm$^{-3}$ is also broadly consistent with the results 
of the recent {\it Bolocam Galactic Plane Survey} of dense molecular clouds (Rosolowsky et al. 2010).
As the gas content of the Galaxy today is lower than in the past due to star formation and feedback, in equation (14) we have 
chosen a normalization for $f_{\rm cloud}$ that is a factor of 10 larger than that derived by Talbot \& Newman for the case 
of the present-day Milky Way, in order to conservatively account for the higher gas fraction in the high redshift progenitors of the Galaxy.
Assuming that the star formation rate (SFR) within a halo is proportional to the mass of dense gas 
(e.g. $n_{\rm gas}$ $\ga$ 30 cm$^{-3}$; Wolfire et al. 2003; Krumholz et al. 2009) in the halo, that the 
mass of dense gas is proportional to the total halo mass, and that the SFR in the Galaxy today 
is of the order of 1 M$_{\odot}$ yr$^{-1}$, this normalization yields a SFR of order 10$^{-2}$ M$_{\odot}$ yr$^{-1}$ for dwarf galaxies in 
$\sim$ 10$^{9}$ M$_{\odot}$ haloes at $z$ $\la$ 10, in broad agreement with the SFRs found in recently high-resolution cosmological 
simulations (Wise \& Cen 2009; Razoumov \& Sommer-Larsen 2010).  
As we will show, the probability for accretion in the 
halo is much higher at high redshift than in the present-day Milky Way, and thus our choice of an elevated volume-filling fraction of 
dense gas compared to the present-day Galaxy does not appreciably affect our results for the overall probability of accretion, even though it is clearly an overestimate at low redshift.


Using the mass accretion history of the main progenitor halo given by equation (13) and the volume-filling fraction $f_{\rm cloud}$ given by equation (14), 
we estimate the number of cold ($c_{\rm s}$ $<$ $v_{\rm max}$), dense clouds $N_{\rm acc}$($z$) with $n_{\rm gas}$ $\ga$ $n_{\rm min}$ (given by equation 11) through 
which a star has passed between redshift $z$ and the present-day as

\begin{equation}
N_{\rm acc} \sim \int^{z}_{0} \frac{f_{\rm cloud}} {t_{\rm orb}} \left|\frac{dt}{dz'}\right| dz' \mbox{\ .}
\end{equation}  
where $t_{\rm orb}$ $\sim$ $r_{\rm ap}$ / $v_{\rm max}$ is the orbital period of the star, and $t$($z$) is age of the Universe at redshift $z$. 
While it is largely uncertain what we should choose for the apogalactic distance of the stellar orbit, for our current purposes
we choose $r_{\rm ap}$ = 0.1 $r_{\rm vir}$, to be consistent with simulations which suggest that Pop~III stars formed in the high-redshift progenitors of the Galaxy
end up in the central $\sim$ 0.1 $r_{\rm vir}$ of the halo at $z$ = 0 
(e.g. Diemand et al. 2005; Scannapieco et al. 2006; Brook et al. 2007; De Lucia \& Helmi 2008; Madau et al. 2008; Salvadori et al. 2010; Gao et al. 2010).
We shall relax this assumption in the next Section, in which we consider the broader question of accretion onto Pop~III stars formed within the full population of 
Milky Way progenitor haloes.

We emphasize that the formula for $f_{\rm cloud}$ that we use here (equation 14) was derived by Talbot \& Newman (1977) as the 
volume-filling fraction of the disk of the Milky Way.  Therefore, our estimate of the number $N_{\rm acc}$ of dense clouds encountered 
implicitly assumes that at every orbit a star passes through the central, star-forming 
regions of its host halo corresponding to the disk of the Galaxy today. As stars may instead orbit completely outside the central regions of the halo 
(i.e. at radii $\ga$ 0.1 $r_{\rm vir}$), in this sense equation (15) is an upper limit for $N_{\rm acc}$.  

Evaluating the integral in equation (15) yields a probability of accretion after redshift $z$ of $p_{\rm acc}$($< z$) $\sim$ 1 - $e^{-N_{\rm acc}}$, 
which is shown in Figure 1.  One trend that is evident is that the probability for accretion is much higher at high redshift.  
There are two reasons for this.  Firstly, the maximum circular velocity $v_{\rm max}$ of the halo is lower at higher redshift, and hence the 
volume-filling factor of gas at densities above $n_{\rm min}$ is higher.  Secondly, the orbital period $t_{\rm orb}$ of stars that we have 
chosen is also lower at high redshift, leading to more frequent passages through the ISM during which accretion may take place.  We also note that there is a probability of only 
$p_{\rm acc}$ $\la$ 0.2 of accretion onto stars formed at $z$ $\la$ 20, and a probability of only $p_{\rm acc}$ $\sim$ 10$^{-5}$ of accretion onto such stars during passages through 
the disk of the Galaxy over the past $\sim$ 10 Gyr, as treated in Frebel et al. (2009).  From this we conclude that, if they have Solar-like winds, 
the majority of any extant low-mass Pop~III stars formed in the main progenitor halo of the Milky Way are likely to exhibit pristine surface compositions today,
having been unaffected by the accretion of metal-enriched gas.

\begin{figure*}
\includegraphics[width=6.5in]{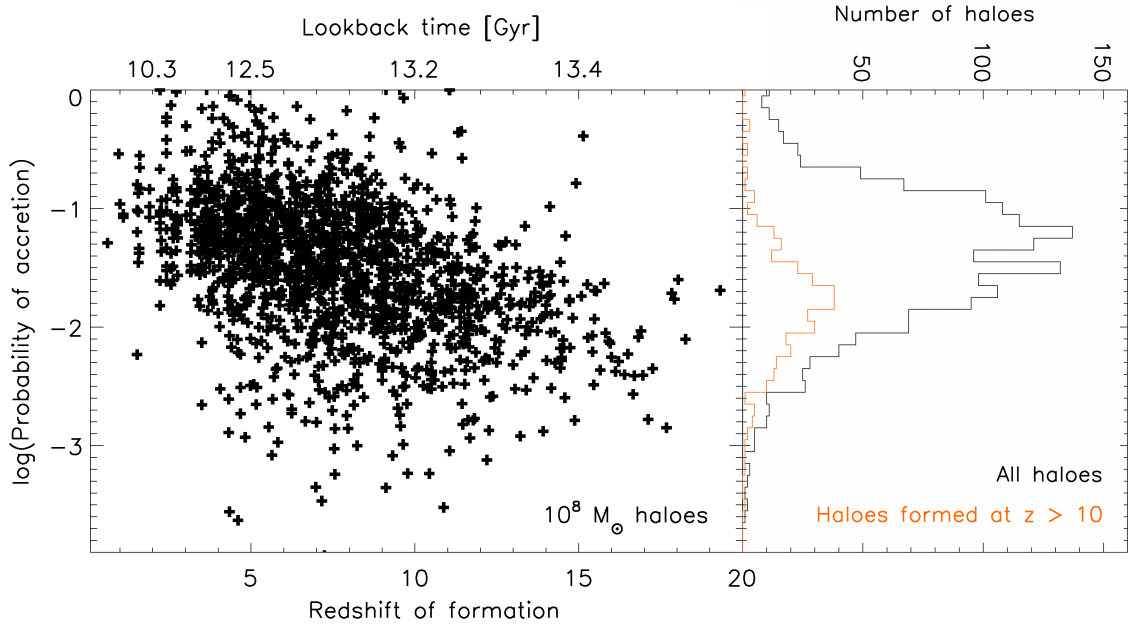}
\caption{The probability of accretion $p_{\rm acc}$ onto low-mass Pop~III stars hosted by the 10$^8$ M$_{\odot}$ progenitor haloes of the DM halo of the Milky Way, as a function 
of the redshift of formation of the haloes ({\it left panel}).  The distribution of these probabilities is shown in the right panel, for all 10$^8$ M$_{\odot}$
haloes ({\it black histogram}) and for the subset of 10$^8$ M$_{\odot}$ haloes formed at $z$ $>$ 10 ({\it red histogram}).  
While accretion is expected to occur in a small subset of haloes hosting Pop~III stars at $z$ $\le$ 10, in general the probability 
of accretion is only $\le$ 0.1.  The probability is almost an order of magnitude lower for Pop~III stars formed at $z$ $>$ 10, at which redshifts the bulk of 
Pop~III stars are likely to have formed prior to widespread chemical enrichment and the onset of reionization.  
}
\end{figure*}

\subsubsection{Accretion within all progenitor haloes}

While the earliest Pop~III stars are most likely to form in the main progenitor halo of the Galaxy, the majority may form 
in other, smaller progenitor haloes.  Here we evaluate the probability of accretion onto low-mass Pop~III stars in these haloes.
For this purpose, we employ the extended Press-Schechter (e.g. Lacey \& Cole 1993) formalism to construct 
a merger tree of the progenitor haloes of a Milky Way-like halo, with a mass of 10$^{12}$ M$_{\odot}$ at $z$ = 0.  In particular, we employ the 
Monte-Carlo method suggested by Somerville \& Kolatt (1999).

As we have shown, in haloes of lower mass, accretion is more likely to occur, due principally to the fact that a star moving at lower velocity
$v_{\rm *}$ $\sim$ $v_{\rm max}$ can accrete gas at a lower density, which in turn has a larger volume-filling factor $f_{\rm cloud}$.  Therefore, the minimum
halo mass for which accretion of metals can take place is a critical parameter in our calculation.  As discussed in Section 4.1, a minimum halo 
mass of 10$^{8}$ M$_{\odot}$ is likely a good estimate (e.g. Greif et al. 2010b), as this is the typical mass of a halo that is able to re-capture the metal-enriched ejecta of the first
Pop~III SNe, which typically occur in the 10$^6$ - 10$^7$ M$_{\odot}$ progenitors of these haloes (e.g. Bromm \& Larson 2004; Trenti \& Stiavelli 2009).

Similar to the case of the main progenitor halo described in Section 4.2.1, we estimate the probability of accretion onto a low-mass Pop~III star 
by integrating equation (15) from $z$ = 0 to the redshift at which a halo with the minimum mass forms.  However, to account for the numerous halo mergers that
alter the orbital parameters of a star, and in particular the apogalactic distance of its orbit, instead of assuming $r_{\rm ap}$ = 0.1 $r_{\rm vir}$
as before, we now use $r_{\rm ap}$ = $N_{\rm rand}$ $r_{\rm vir}$, where $N_{\rm rand}$ is a random number between zero and unity that is updated after each merger. 
This yields the orbital period of a star in a given halo
$t_{\rm orb}$ = $N_{\rm rand}$ $r_{\rm vir}$ / $v_{\rm max}$.  Also, as in Komiya et al. (2010), we account for the delay in a star being incorporated into its 
new host halo after a merger, which is taken to be the timescale for dynamical friction given by Springel et al. (2001).  
Similar to the case of accretion onto Pop~III stars with no winds considered by Komiya et al., we find that the accounting for this delay results in 
somewhat higher accretion probabilities, especially for stars formed in haloes at relatively low $z$ that undergo mergers with much more massive haloes (see discussion below).  

Figure 2 shows the probability that low-mass Pop~III stars within haloes with mass $\ge$ 10$^8$ M$_{\odot}$ accrete metal-enriched gas during the 
assembly of the Galaxy.  The left panel shows the probability of accretion onto stars hosted by all of the 10$^8$ M$_{\odot}$ progenitor haloes, as a function
of the redshift at which they form.  The right panel shows the distribution of these probabilities; assuming that each 10$^8$ M$_{\odot}$ halo hosts the same number
of low-mass Pop~III stars, this can be translated as the distribution of probabilities of accretion onto all Pop~III stars in the halo of the Galaxy today. 
As shown by the black histogram in the right panel, the probability of accretion is generally low, with the stars in most haloes having a probability of $\la$ 0.1
to accrete metal-enriched gas.  We note that, although we have simply assumed that the stars have random orbits in their host halos, 
this conclusion holds for different random realizations of these orbits (i.e. different $N_{\rm rand}$).  

Interestingly, as shown in the left panel of Fig. 2, the haloes for which accretion is most likely are those which form
at relatively low redshift, $z$ $\la$ 10.  This is a result of the fact that such late-forming haloes also grow slowly, and 
so there is more time available for accretion of relatively low-density gas in these haloes compared to those that grow more rapidly at higher redshift.
Another important effect is that a portion of the late-forming 10$^8$ M$_{\odot}$ haloes also undergo mergers with much more massive haloes than at higher $z$.
As the timescale for dynamical friction in these mergers can be of the order of Gyr, the stars can orbit within 
the haloes of lower mass (and hence lower $v_{\rm max}$) for much longer times than they would at higher $z$.  In turn, this results 
in considerably higher accretion probabilities for some stars formed at relatively low $z$; however, the effect of dynamical
friction on the overall probability distribution, shown in the right panel of Fig. 2, is not large, as also found by Komiya et al. (2010)
in a similar calculation.

However, it is important to note that the fraction of such haloes at low redshift that actually host Pop~III stars may be relatively low.  This is 
due to feedback effects from stars formed in neighboring haloes, such as metal enrichment (e.g. Maio et al. 2010; Greif et al. 2010b) and photoevaporation
(e.g. Shapiro et al. 2004; Whalen et al. 2008b), which will act to diminish the formation 
rate of Pop~III stars (see also Tsujimoto et al. 1999; Hernandez \& Ferrara 2001).  Even neglecting the suppression of Pop~III star formation due to metal enrichment 
and photoionization by massive stars, the suppression due to the photodissociation of molecules (e.g. Haiman et al. 1997) is likely to be significant on its own. 
Assuming a physical number density of atomic-cooling haloes of $\sim$ 0.03 kpc$^{-3}$ at $z$ $\la$ 20, as found from simulations of the formation of 
Milky Way-like haloes (Gao et al. 2010), either massive stars (e.g. Yoshida et al. 2003) or rapidly 
accreting black holes (e.g. Greif et al. 2008; Johnson et al. 2010) within these galaxies would produce a background molecule-dissociating flux 
of the order of $J_{\rm 21}$ $\sim$ 10$^{-2}$ (i.e. 10$^{-2}$ $\times$ 10$^{-21}$ erg s$^{-1}$ cm$^{-2}$ Hz$^{-1}$ sr$^{-1}$), high enough to suppress the 
rate of formation of Pop III stars in minihaloes (e.g. O'Shea \& Norman 2008).  
Furthermore, the probabilities of accretion onto Pop~III stars forming during and after reionization (e.g. at $z$ $\la$ 10) (Johnson 2010) 
are upper limits, since the 10$^8$ M$_{\odot}$ haloes in which they form will have a substantial amount of the gas photoevaporated out of the halo and 
hence not available for accretion (see e.g. Thoul \& Weinberg 1996; Gnedin 2000; Dijkstra et al. 2004).  

To evaluate the impact that external chemical and radiative feedback may have on the probability of accretion onto extant 
low-mass Pop~III stars, the red histogram in the right panel of Fig. 2 shows the distribution of accretion probabilities in 
the subset of haloes formed at $z$ $>$ 10.  
The distribution of probabilities for this subset of haloes peaks at somewhat lower values than that of the overall population, with a typical probability 
for accretion onto a star within these haloes of the order of 10$^{-2}$.  Therefore, we conclude that if chemical enrichment is rapid or radiative feedback is strong
during the assembly of the Milky Way, then the accretion of metal-enriched gas onto a given low-mass Pop~III star is quite unlikely.  

To account for the possibility that the metal-enriched SNe ejecta of the first stars can be contained in somewhat less massive haloes (see e.g. Kitayama \& Yoshida 2005; 
Whalen et al. 2008a), we have also carried out the same exercise with a minimum halo mass for accretion of 10$^7$ M$_{\odot}$.
In this case, we find that the probability distribution for all such haloes peaks at $\sim$ 0.2, and that that for the subset forming at $z$ $>$ 10 peaks at $\sim$ 0.1.  
While the probability for accretion in this case is considerably higher, we can conclude that even in this case the majority of low-mass Pop~III stars are not likely 
to have accreted metal-enriched gas.  

\section{Implications and Conclusions}
We have carried out an analysis of the likelihood that low-mass Pop~III stars with primordial surface compositions 
could be found in the present-day Milky Way, under the assumption that such stars have Solar-like stellar winds 
that prevent the accretion of metal-enriched material from the ISM.  

We have found that if low-mass Pop~III stars are kicked out of the central $\sim$ 10 pc of the haloes in which they form, 
either via dynamical interactions with more massive stars or from violent relaxation during an early merger, 
then the accretion of gas onto the stars can be prevented, at least  
until the star is incorporated into a larger halo in which dense molecular clouds are formed outside of the central $\sim$ 10 pc (e.g. Pawlik et al. 2010).  
To evaluate the likelihood that a star accretes gas during the passage through dense molecular clouds at later stages
in the hierarchical assembly of the Galaxy, we have modeled the dynamics of Pop~III stars in their host dark matter haloes along with the density
distribution of gas within these haloes.  Overall, we find that the probability of enrichment via accretion is $\la$ 0.1,
and is even lower for Pop~III stars formed at $z$ $\ga$ 10, at which times the majority of primordial star formation may have occurred (e.g. Maio et al. 2010).

The primary uncertainty affecting our calculations is in the volume-filling fraction $f_{\rm cloud}$ of dense clouds in the progenitor 
haloes of the Milky Way.  As discussed in Section 4.2.1, we have made the simple, conservative choice of a constant volume-filling fraction.  While this 
is likely an overestimate at low redshift, it does yield broad agreement with cosmological simulations of star formation in dwarf galaxies 
in haloes of mass $\ga$ 10$^8$ M$_{\odot}$ at high redshift (i.e. $z$ $>$ 2), which we have shown are the most likely sites for accretion onto low-mass Pop~III stars.  
We note, however, that existing observations of relatively massive 
galaxies (i.e. within haloes of mass $M_{\rm h}$ $\ga$ 10$^{11}$ M$_{\odot}$) at $z$ $\la$ 2 suggest a steep increase of the 
star formation rate, which is likely to be proportional to the amount of dense gas, with redshift (e.g. Bouch{\' e} et al. 2010).  It is possible that 
such a redshift dependence, which we have not sought to model, extends to the less massive progenitors of the Milky Way at higher redshift.
More precise calculations of the probability of accretion onto low-mass stars will have to await observations, for instance by the 
{\it Atacama Large Millimeter Array} (e.g. Combes 2010), which provide stronger constaints on the properties of the dense gas in high redshift galaxies.    

We emphasize that the low probability we find for accretion onto low-mass Pop~III stars is limited to the case that such stars have weak, Solar-like stellar winds.  
As previous authors have shown, the amount of metal enrichment via accretion can be much higher for stars with no winds (Frebel et al. 2009; Komiya et al. 2010). 
Importantly, unless low-mass Pop~III stars do have winds, such enrichment via accretion is likely to be an inevitability and the likelihood of 
finding truly pristine Pop~III stars in the Galaxy today is practically zero.

Our results thus suggest that it may be possible to test, unequivocally, theoretical predictions of whether or not low-mass Pop~III stars formed 
in the early Universe, as we have found that any such stars are likely to still exhibit primordial surface compositions in the 
Galaxy today.  The assessment of these theories may ultimately require that such pristine stars be found (see e.g. Weiss et al. 2000) in large surveys 
of metal-poor stars, such as the {\it Sloan Extension for Galactic Understanding and Exploration} (e.g. Yanny et al. 2009), 
the {\it Radial Velocity Experiment} (e.g. Fulbright et al. 2010), the {\it Apache Point Observatory Galactic Evolution Experiment} 
(e.g. Allende Prieto et al. 2008), the {\it Large Sky Area Multi-Object Fiber Spectroscopic Telescope} project (e.g. Zhao et al. 2006), 
and the HERMES\footnote{http://www.aao.gov.au/AAO/HERMES/} project.  
If, however, they are not discovered, then it may be for one of two reasons: either low-mass Pop~III stars do not have sufficiently 
strong winds to prevent enrichment via accretion, or long-lived Pop~III stars were not formed in sufficiently high numbers in the progenitor 
haloes of the Milky Way (see e.g. Madau et al. 2008) for them to be found in observational campaigns.

\section*{Acknowledgements}
JLJ is grateful to Thomas Greif for sharing his results on the kinematics of 
low-mass Pop~III stars in minihaloes ahead of their publication, as well as for comments on an 
early draft of this work. 
The authors would like to thank Fabrice Durier, Volker Gaibler, and Claudio Dalla Vecchia
for valuable feedback and discussion, as well as to the anonymous reviewer for a helpful report.


\end{document}